\documentclass[showpacs,twocolumn,superscriptaddress,aps]{revtex4-1}

\usepackage{graphicx}
\usepackage{amsmath,amssymb,bm}
\usepackage{epstopdf}

\newcommand{\iChEM}{{\it i}{\rm ChEM}}

\newcommand{\tS}{\text{\tiny S}}
\newcommand{\B}{\text{\scriptsize res}}

\newcommand{\SB}{\text{\scriptsize sys-res}}
\newcommand{\T}{{\rm total}}
\newcommand{\dg}{\dagger}

\newcommand{\up}{\uparrow}
\newcommand{\down}{\downarrow}

\newcommand{\w}{\omega}

\newcommand{\AF}{\text{\tiny AF}}
\newcommand{\K}{\text{\tiny K}}

\newcommand{\be}{\begin{equation}}
\newcommand{\ee}{\end{equation}}
\newcommand{\bea}{\begin{eqnarray}}
\newcommand{\eea}{\end{eqnarray}}
\newcommand{\bsube}{\begin{subequations}}
\newcommand{\esube}{\end{subequations}}

\newcommand{\Fig}[1]{Fig.\,\ref{#1}}

\begin{document}


\title{Ferromagnetic Phase in Nonequilibrium Quantum Dots}

\author{WenJie Hou}
 \affiliation{Department of Physics, Renmin University of China, Beijing 100872, China}

\author{YuanDong Wang}
 \affiliation{Department of Physics, Renmin University of China, Beijing 100872, China}

\author{JianHua Wei}\email{wjh@ruc.edu.cn}
 \affiliation{Department of Physics, Renmin University of China, Beijing 100872, China}

\author{YiJing Yan}
 \affiliation{Hefei National Laboratory for Physical Sciences at the Microscale and
 \iChEM\ (Collaborative Innovation Center of Chemistry for Energy Materials),
 University of Science and Technology of China, Hefei, Anhui 230026, China}

\date{\today}

\begin{abstract}
 By nonperturbatively solving the nonequilibrium Anderson two-impurity model with the hierarchical equations of motion approach, we report a robust ferromagnetic (FM) phase in series-coupled double quantum dots, which can suppress the antiferromagnetic (AFM) phase and dominate the phase diagram at finite bias and detuning energy in the strongly correlated limit. The FM exchange interaction origins from the passive parallel spin arrangement caused by the Pauli exclusion principle during the electrons transport. At very low temperature, the Kondo screening of the magnetic moment in the FM phase induces some nonequilibrium Kondo effects in magnetic susceptibility, spectral functions and current. In the weakly correlated limit, the AFM phase is found still stable, therefore, a magnetic-field-free internal control of spin states can be expected through the continuous FM--AFM phase transition.
\end{abstract}

\pacs{}
\maketitle

\begin{figure*} [tbph] \centering
\includegraphics[width=5.0in]{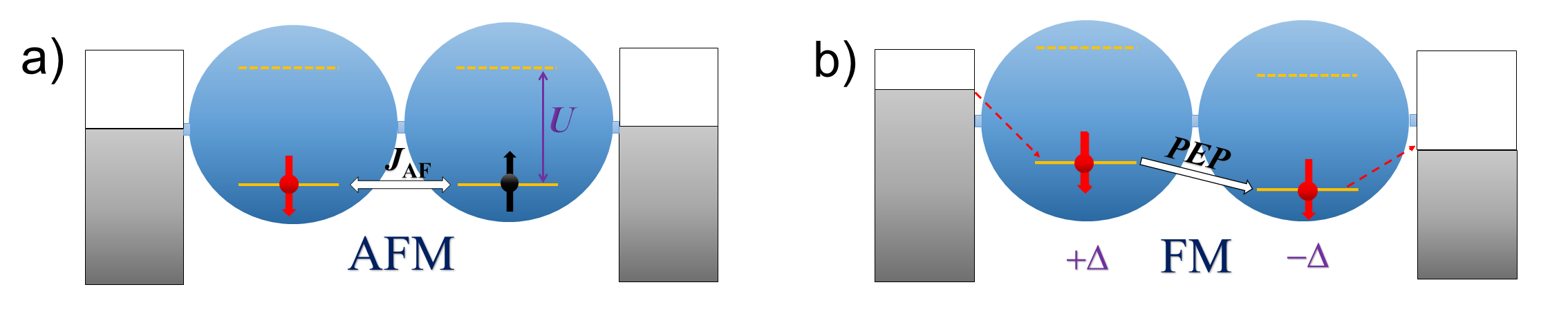}
\includegraphics[width=6.0in]{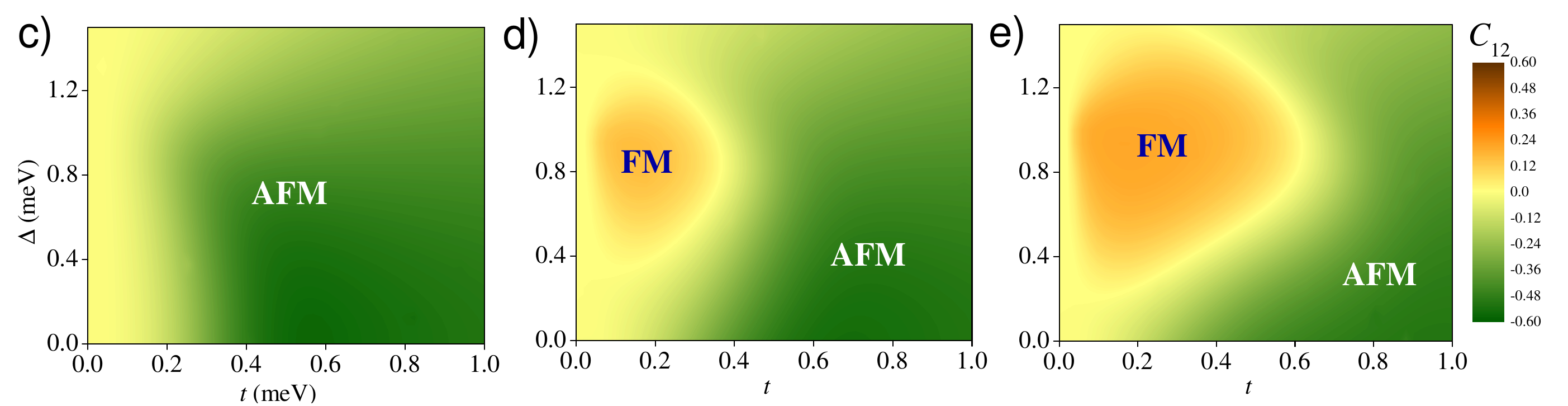}
\caption{(Color online). (a) Schematic diagram of antiferromagnetic (AFM) state in series-coupled double quantum dots (SDQDs) under equilibrium conditions. $U$ is the on-dot electron-electron ($e-e$) interaction. $J_{\AF}$ is the strength of AFM exchange interaction. (b) Schematic diagram of ferromagnetic (FM) state in SDQDs under nonequilibrium conditions at finite bias $V$ and detuning energy $2\Delta$. PEP denotes the Pauli exclusion principle during the electrons transport. (c)--(e) Magnetic phase diagrams of SDQDs in the $\Delta-t$ ($t$ being the inter-dot coupling) plane by showing the spin-spin correlation function $C_{12}\equiv\langle \vec{S}_{1}\cdot \vec{S}_{2} \rangle-\langle \vec{S}_{1}\rangle\cdot\langle \vec{S}_{2}\rangle$ at various bias, (c) $V=0$; (d) $V=0.5$ mV and (e) $V=1.0$ mV.}\label{fig1}
\end{figure*}

The ferromagnetism intrinsically origins from the spin-independent Coulomb interaction and the Pauli exclusion principle (PEP), as initially proposed by Heisenberg \cite{Hei28619}.   The Hubbard model \cite{Hub63238}, which includes both two elements with on-site electron-electron ($e-e$) interaction $U$, is regarded as the minimal model for ferromagnetic (FM) states. Unfortunately, it has not been well addressed whether the Hubbard model has a general FM phase, except under some special conditions\cite{Nag66392,Lie891201,Tas954678,Bat02187203}. The Hartree-Fock approximation once predicted an itinerant Stoner-like FM phase \cite{Sto331018}, but we now know that the mean-field theory deduces incorrect results and the FM region has been overestimated  \cite{Nag66392}. Besides the Hubbard model, the Anderson (multi)-impurity model \cite{And6141} may act as another minimal model for magnetic phase in a bottom-up fashion, with the advantage of implementation simplicity in quantum dots (QDs). For example, the antiferromagnetic (AFM) correlation $J_{\AF}$ due to nearest-neighbour electron hopping or tunneling $t$ ($J_{\AF}\sim 4t^2/U$) has been well understood experimentally in series-coupled double QDs (SDQDs) [see \Fig{fig1}(a)] \cite{Cha09096501}.  Theoretically, $J_{\AF}$ is responsible for the AFM ground state at half filling in the Hubbard model, while it induces the spin singlet competing with the Kondo singlet at temperature $T<T_{\K}$ ($T_{\K}$ being the Kondo temperature) in the Anderson two-impurity model \cite{Cha09096501,Jon88843,Aff921046,Che04176801,Li12266403}.

Does there exist a FM phase in the Anderson two-impurity model or in SDQDs?  That issue may help to understand Heisenberg's original idea and to determine the FM phase in various strongly correlated models.  Please be noted that the sign-indefinite Ruderman-Kittel-Kasuya-Yosida (RKKY) magnetic order, whose implementation must through a third mediated dot in experiments\cite{Cra04565}, is not our concern here. What we are seeking is a stable FM phase strong enough to compete with the AFM one in SDQDs, which has not been explicitly determined yet in the phase diagrams of SDQDs \cite{Cha09096501} and other two-impurity systems \cite{Bor11901}.

The FM phase in SDQDs also has great application potential in solid-state quantum computing.  QDs-based spin qubit is one of the most possible physical realization of scalable qubit put forward so far, which has been extensively studied in last two decades \cite{Han071217,Zwa13961} since its original proposal in SDQDs \cite{Los98120}. It has the advantages of fast operation and long coherence times but the disadvantage of seriously dependence on magnetic fields. The technical difficulties caused by magnetic fields are transparent:  ($i$) The localized oscillating magnetic fields required in qubit or quantum gate manipulation are very hard to realize in practice; ($ii$) The Zeeman energy is an inefficient way to control spin states; and ($iii$) The magnetic fields are incompatible with present large-scale integrated circuit.  If a stable FM phase in SDQDs does exist, these difficulties may be overcome by possible magnetic-field-free manipulations.

In the present work, by nonperturbatively solving the Anderson two-impurity model, we will firstly verify no FM phase in the range of parameters investigated under the equilibrium condition in SDQDs. Then, we will report a robust FM phase under nonequilibrium conditions at finite bias and detuning energy, which are strong enough to suppress the AFM phase in the strongly correlated limit ($t\ll U$). We will demonstrate that the FM exchange interaction origins from the passive parallel spin arrangement caused by the PEP during the electrons transport [see \Fig{fig1}(b)].  The FM phase is the effect of PEP on magnetic properties, beyond the current collapse in DQDs, another effect of PEP called Pauli spin blockade \cite{Ono021313,Mur07035432,Hou17224304}. At large $t$, the AFM phase keeps stable, which defines a tunnel-barrier control of spin states through
the FM--AFM transition in SDQDs, similar to the initial proposal in Ref.~[\onlinecite{Los98120}] but no magnetic field (or auxiliary FM-dots) needed any more.

The SDQDs we study here can be described by the nonequilibrium Anderson two-impurity model.  The total Hamiltonian reads $H_{\T}=H_{\tS}+H_{\B}+H_{\SB}$, where the isolated QD part is
\begin{align}\label{HS}
H_{\tS}=\sum_{i,s} \epsilon_{i,s}\hat{c}^\dag_{i,s}\hat{c}_{i,s} + \frac{U}{2}\sum_{i,s}\hat{n}_{i,s}\hat{n}_{i\bar{s}}+t\sum_{s}(\hat{c}^\dag_{1,s}\hat{c}_{2,s}+\text{h.c.}),
\end{align}
here $\hat{c}_{i,s}^\dag$ ($\hat{c}_{i,s}$) is the operator that creates (annihilates) an $s$-spin ($s=\up,\down$) electron with energy $\epsilon_{i,s}$ in the dot $i$ ($i=1,2$). $\hat{n}_{i,s}=\hat{c}^\dag_{i,s}\hat{c}_{i,s}$ corresponds to the $s$-spin electron number operator of dot $i$. As mentioned above, $U$ ($U=U_1=U_2$) is the on-dot Coulomb interaction between $s$- and $\bar{s}$-spin electrons ($\bar{s}$ being the opposite spin of $s$), and $t$ is the interdot coupling strength.

The Hamiltonians of reservoirs   are $H_{\B}=\sum_{\alpha ks}
(\varepsilon_{\alpha ks}+\mu_{\alpha})\hat{c}^\dg_{\alpha ks}\hat{c}_{\alpha ks}$, $\alpha={\rm L,R}$,
under the bias  $V=(\mu_{\rm L}-\mu_{\rm R})/e$, where $\hat c^{\dg}_{\alpha ks}$ ($\hat c_{\alpha ks}$)
denotes the creation (annihilation) operator of an electron in the $s$-spin state in the $\alpha$-reservoir with wave vector $k$.
We set the Fermi energy $E_{\rm F}=\mu^{\rm eq}_{\rm L}=\mu^{\rm eq}_{\rm R}=0$ at equilibrium and $\mu_{\rm L}/e=-\mu_{\rm R}/e=V/2$ at nonequilibrium. The system-reservoir coupling is $H_{\SB}=\sum_{\alpha kis} t_{\alpha kis}\hat c^{\dg}_{is}\hat c_{\alpha ks}+{\rm h.c.}$. The hybridization function is assumed to be a Lorentzian form $J_{\alpha is}(\w)=\pi\sum_{k} t_{\alpha kis}t^{\ast}_{\alpha kis} \delta(\omega-\varepsilon_{\alpha ks})
= \Gamma W^2/(\omega^2+W^2)$.

\begin{figure*} 
\includegraphics[width=6.0in]{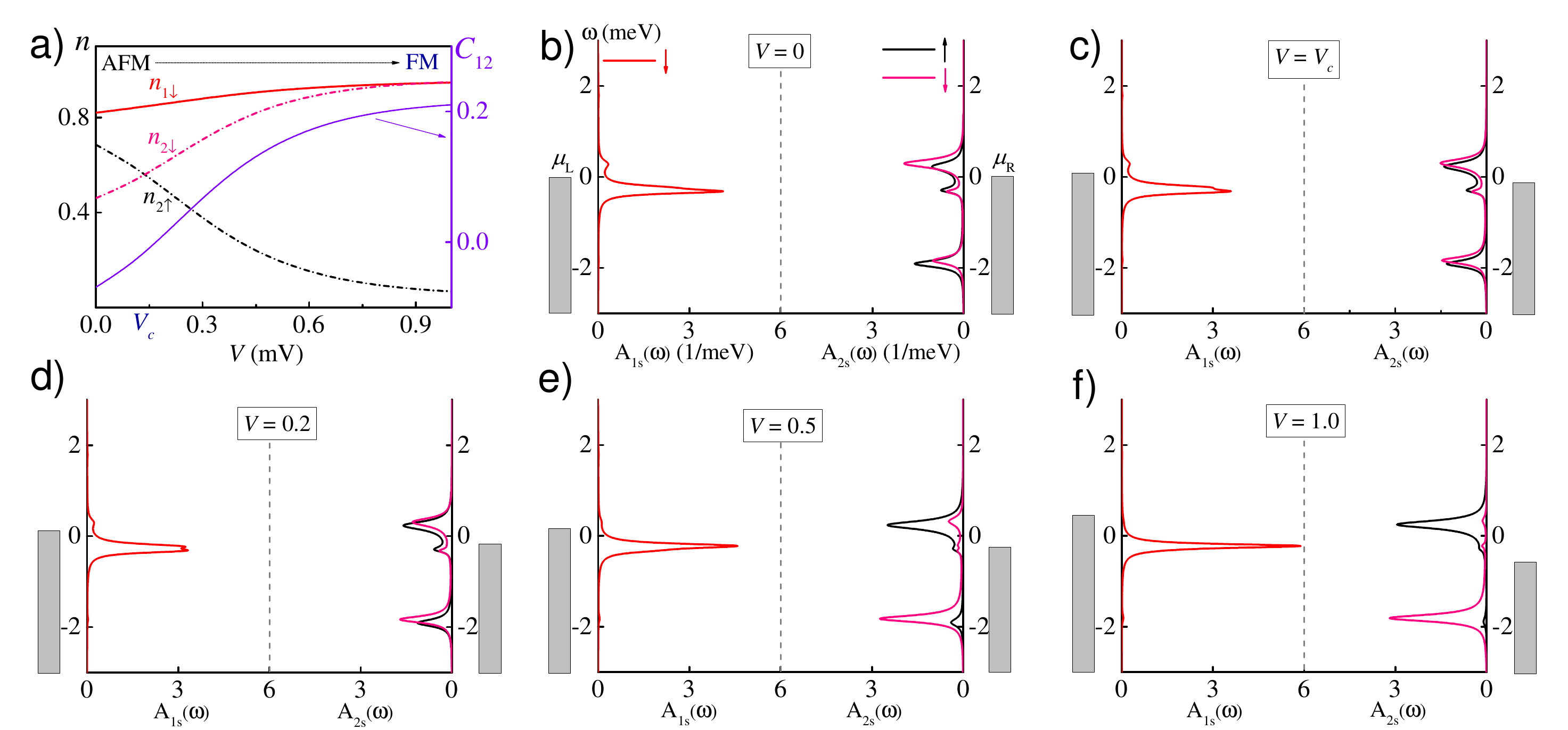}
\caption{(Color online). In the case of spin non-degeneracy in QD1 at $t=0.2$ meV and $\Delta=0.75$ meV. (a) The dependence of
$n_{1\down}$, $n_{2\up}$, $n_{2\down}$ and $C_{12}$ on $V$. $V_c=0.14$ mV is the AFM--FM phase crossover point. (b)--(f)  The spectral functions $A_{1\down}(\w)$, $A_{2\up}(\w)$ and $A_{2\down}(\w)$ at (b) $V=0$; (c) $V=V_c$; (d) $V=0.2$ mV; (e) $V=0.5$ mV;  and (f) $V=1.0$ mV. The unit of $V$ in the figure is mV.}
\label{fig2}
\end{figure*}

We adopt the hierarchical equations of motion (HEOM) approach \cite{Jin08234703,Li12266403} to numerically solve the nonequilibrium Anderson two-impurity model in a nonperturbative fashion. The HEOM can achieve the same level of accuracy
as the latest high-level numerical renormalization group (NRG) \cite{Wil75773} for both static and dynamical quantities under equilibrium conditions \cite{Li12266403}. Under nonequilibrium conditions, the HEOM has many advantages above other approaches in the prediction of dynamical properties \cite{Che15033009,Che17155417,Hou172486,Hou17224304}. The HEOM formalism is in principle exact and applicable to arbitrary electronic systems, including Coulomb interactions, under the influence of arbitrary time-dependent bias voltage and external fields  \cite{Jin08234703,Li12266403, Che15033009,Che17155417,Hou172486,Hou17224304,Zhe09164708,Zhe121129,Jin15234108}. The details of the HEOM formalism and the derivation of physical quantities are supplied in Refs.~[\onlinecite{Jin08234703}],~[\onlinecite{Li12266403}] and [\onlinecite{YeWIREs}].

The parameters in our calculations are chosen as follows: the on-dot $e-e$ interaction $U=2.0$ meV; the singly occupied energy level $\epsilon_{1\down}=\epsilon_{1\up}=-1.0\;{\rm meV}+\Delta$ and $\epsilon_{2\down}=\epsilon_{2\up}=-1.0\;{\rm meV}-\Delta$, where the detuning energy $2\Delta$ can be finely regulated by gate voltages in experiments; the temperature $T=0.1$ meV unless otherwise noted; the effective bandwidth of the reservoirs $W_{\rm L}=W_{\rm R}=W=4.0$ meV and the reservoir-dot coupling strength $\Gamma_{\rm L}=\Gamma_{\rm R}=\Gamma=0.1$ meV. The inter-dot coupling $t$, bias of voltage $V$ and detuning energy $\Delta$ are three main variables in our calculations.

In order to figure out whether there exists a FM state,  we calculate the spin-spin correlation function between QD1 and 2,
\begin{equation}\label{C12}
 C_{12}\equiv\langle \vec{S}_{1}\cdot \vec{S}_{2} \rangle-\langle \vec{S}_{1}\rangle\cdot\langle \vec{S}_{2}\rangle,
 \end{equation}
where $\vec{S}_{i}$ is the quantum spin operator at dot $i$. In \Fig{fig1}(c)--(e), we depict the phase diagram at bias $V=0$, $0.5$ and $1.0$ mV, characterized by the sign and value of $C_{12}$ in the $\Delta-t$ plane.  Under the equilibrium condition, as shown in \Fig{fig1}(c), the sign of $C_{12}$ keeps always negative, which indicates a single AFM phase independent of
$t$ ($t>0$)  and $\Delta$. It is understandable.  From the second-order perturbation, one can obtain $J_{\AF}\sim 4t^2U/[U^2-(2\Delta)^2]$ at finite $\Delta$, seeming a negative $J_{\AF}$ included. However, the condition for that equation ($t\ll U$ and $\Delta<U/2$) makes  $J_{\AF}<0$ impossible, even under nonequilibrium conditions. Thus, the following FM phase can not result from this mechanism. As shown in \Fig{fig1}(c), with increasing $t$, $C_{12}$ positively increases, and finally an AFM QD-molecule forms in the large $t$ limit \cite{Wu14}, as an analogue of hydrogen molecule.

When a positive bias applied, as shown in \Fig{fig1}(d) and (e), our results reveal a FM phase appearing in the region of $0<t\ll U$ and $0.2U<\Delta<0.7U$. In view of the phase changes from \Fig{fig1}(d) to (e), the FM phase can be seen as growing from the AFM background at finite bias. The FM--AFM phase boundary (where $C_{12}$ changing its sign) seems quite smooth with no abrupt phase transition occurring, instead, a continuous crossover behaviour is clearly visible. With increasing bias, the area of FM phase is enlarged and the strength of exchange interaction enhanced, as $C_{12}$ positively increases. In the strongly correlated limit ($0<t\ll U$), the FM phase can well suppress the AFM one and dominate the phase diagram at finite $V$ and $\Delta$, as shown in \Fig{fig1}(e). However, the AFM molecular state will survive at large $t$ and very small $\Delta$, which respectively determine the right and bottom boundary of FM phase. If $\Delta$ is too large to destroy the single occupation of any dot, $C_{12}$ will decrease to zero rapidly, which determines the upper boundary. The left boundary is naturally at $t\sim0$. As a comprehensive result, the FM phase forms a closed irregular circle area in the phase diagram, as shown in \Fig{fig1}(d) and (e).

\begin{figure*} 
\includegraphics[width=6.0in]{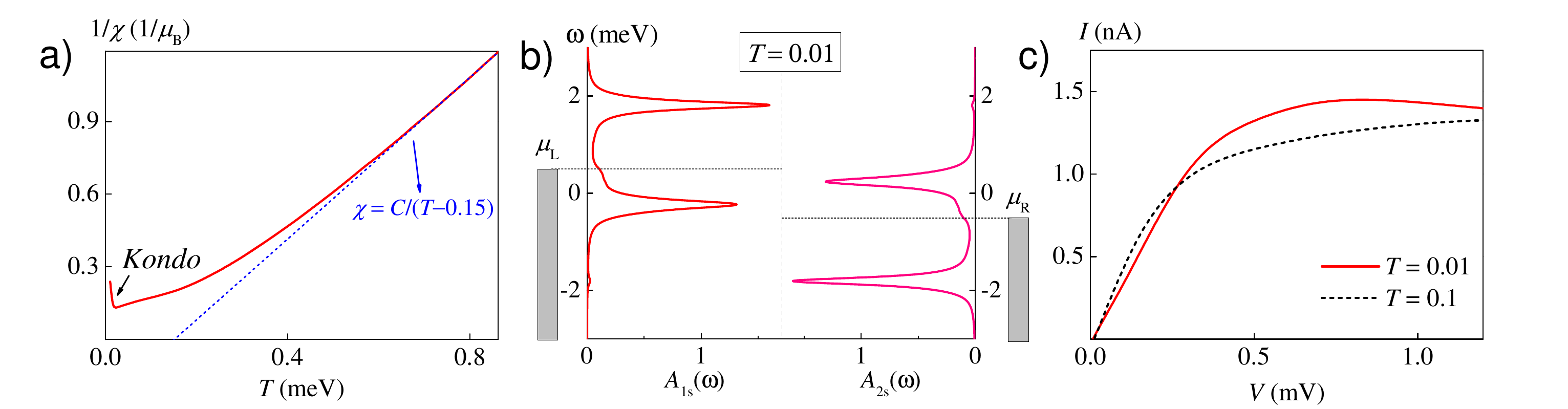}
\caption{(Color online). In the case of spin degeneracy in both QDs at $V=1.0$ mV, $t=0.2$ meV and $\Delta=0.75$ meV. (a) The dependence of the inverse of magnetic susceptibility $1/\chi$ on temperature $T$. The dash line is the fitting of the Curie-Weiss law at high temperature. A Kondo screening effect is shown at $T<0.02$ meV. (b) The spectral functions $A_{is}(\w)$s at temperature below ($T=0.01$ meV) the Kondo temperature. The current-voltage ($I-V$) curves at temperature below ($T=0.01$ meV, the solid line) and above ($T=0.1$ meV, the dashed line) the Kondo temperature.  The unit of $T$ in the figure is meV.}
\label{fig3}
\end{figure*}

In order to better understand the details of the AFM--FM transition, we theoretically lift the spin degeneracy in QD1 by applying a local  magnetic field $B_{1}$, with its direction paralleling to $\down$-spins.  $B_{1}$ is chosen to be strong enough to push  $\epsilon_{1\up}$ much higher than $\mu_{\rm L}$ but left $\epsilon_{1\down}=-1.0 {\rm meV}+\Delta$, which can be achieved by simultaneously adjusting the gate voltage on QD1. By fixing $t=0.2$ meV and $\Delta=0.75$ meV, we calculate both static and dynamical quantities as functions of $V$ and summarize the results in \Fig{fig2}, where \Fig{fig2}(a) depicts some typical static quantities ($n_{1\down}$, $n_{2\up}$, $n_{2\down}$ and $C_{12}$) and \Fig{fig2}(b)--(e) show the spectral functions [$A_{1\down}(\w)$, $A_{2\up}(\w)$ and $A_{2\down}(\w)$] at $V=0$, 0.14 ($V_c$, AFM--FM phase crossover point), 0.2, 0.5, 1.0 mV, respectively.  As a starting point, the AFM phase  at $V=0$ is clearly shown in \Fig{fig2}(a), where the magnetic moments $m_1\equiv n_{1\up}-n_{1\down}\approx -n_{1\down}<0$ and $m_2\equiv n_{2\up}-n_{2\down}>0$.  Accordingly, the degeneracy of $A_{2\up}(\w)$ and $A_{2\down}(\w)$ is lifted due to the AFM exchange interaction $J_{\AF}$,  as shown in \Fig{fig2}(b), where the singly-occupation transition peak of $A_{2\up}(\w)$ is higher than that of $A_{2\down}(\w)$.

Under nonequilibrium conditions, $\down$-spin electrons irreversibly flow from L- to R-reservoir through interdot tunneling. During the transport process, the PEP  affects both electrical \cite{Ono021313,Mur07035432,Hou17224304} and magnetic properties, of which the latter is our focus here. In \Fig{fig2}(a), the continuous crossover from AFM to FM phase is shown in detail. With increasing $V$, $n_{2\up}$ gradually decreases while  $n_{2\down}$ increases, thus $m_2$ positively decreases. At $V\sim 0.14$ mV, $n_{2\up}=n_{2\down}\Rightarrow m_2\sim 0$. As a consequence, $C_{12}\sim 0$, which defines an AFM--FM phase crossover point, $V_c$, as shown in \Fig{fig2}(a). By checking the spectral functions, we find
 the singly-occupation transition peak of $A_{2\up}(\w)$ almost overlaps with that of $A_{2\down}(\w)$ at $V=V_c$ with a little splitting [see \Fig{fig2}(c)]. With further increasing $V$ at $V>V_c$, $m_2$ becomes to negatively increase and $C_{12}$ positively increase, as shown in \Fig{fig2}(a), thus the FM phase is gradually enhanced.  At $V\sim 0.9$ mV,  both $m_2$ and $C_{12}$ reach their saturation values of  0.9  and 0.21, respectively. The continuous increase of $C_{12}$ with a smooth sign change indicates the competition between AFM and FM phases is far from intense.

Fundamentally, finite bias injects $\down$-spin electrons from L-reservoir into QD1, followed by interdot tunneling to QD2. In the next step, the PEP prohibits the double occupation of two $\down$-spin electrons, and electrons can only flow out through off-resonance cotunneling \cite{Ave902446} or many-body tunneling\cite{Hou172486} into R-reservoir, both of which produce small current. As shown in \Fig{fig1}(b), for electrons in QD2, increasing $V$ and/or $\Delta$ will enhance their inflowing probability and meanwhile decrease their off-resonance outflowing probability. When the former becomes much larger than the latter at $V>V_c$ and $\Delta>0.2U$, $\down$-spin electrons will accumulate within QD2, which induces a positive to negative sign change of $m_2$.  As a consequence, the exchange of $\vec{S}_{1}$ and $\vec{S}_{2}$ produces a FM order characterized by $C_{12}>0$. The spectral functions shown in \Fig{fig2}(d) at $V=0.2$ mV verifies this FM correlation (although still weak) , where the singly-occupation transition peak of $A_{2\up}(\w)$ becomes lower than $A_{2\down}(\w)$.

With further increasing $V$, the FM exchange interaction becomes stronger. In spectral functions, this trend is represented by the gradually increasing of the singly-occupation transition peak of $A_{2\down}(\w)$ and decreasing of that of $A_{2\up}(\w)$ [see \Fig{fig2} (e)]. At $V> 0.9$ mV, the former reaches its maximum value and the latter almost disappears, as shown in \Fig{fig2} (f). By summarizing \Fig{fig2}(a)-(f), one can see that the FM phase in SDQDs origins from the passive parallel spin arrangement caused by the PEP during the electrons transport in the presence of $e-e$ interactions. That mechanism is universal, which should play roles in other strongly correlated models including the Hubbard model.

We are now on the position to elucidate the temperature effect, especially the low temperature properties of the FM phase. In what follows, we  recover the spin degeneracy in QD1 and fix $V=1.0$ mV, $t=0.2$ meV and $\Delta=0.75$ meV. The dependence of the inverse of magnetic susceptibility $1/\chi$ on temperature $T$ is depicted in \Fig{fig3}(a), which shows an unambiguous Curie-Weiss behaviour at high temperature, $\chi=C/(T-T_c)$, with a fitted Curie point $T_c\sim $0.15 meV ($\sim 1.75$ K). We also find a upward deviation at very low temperature $T<0.02$ meV, resulting from the Kondo screening of the FM phase at $T<T_{\K}$.  Under equilibrium conditions, this kind of $S=1$ Kondo screening induces a `singular Fermi liquid state' \cite{Aff90517,Col03220405,Pan17025601}. Here, some nonequilibrium Kondo features are expected.

The present HEOM approach can not directly determine $T_{\K}$ as NRG does, but it can easily obtain spectral functions and current at sufficient low temperature to elucidate nonequilibrium Kondo characteristics.  The HEOM results of $A_{is}(\w)$s  and current-voltage ($I-V$) curve at $T=0.01$ meV are respectively shown in \Fig{fig3}(c) and (d), where the $I-V$ curve at $T=0.1$ meV ($T>T_{\K}$) is also shown for comparison. As shown in \Fig{fig3}(b), one small Kondo peak is developed at $\w=\mu_{\rm L}$ in $A_{1s}(\w)$, and another developed at $\w=\mu_{\rm R}$ in $A_{2s}(\w)$. It can be seen as the DQD extension of the bias-induced Kondo peak splitting in single QDs \cite{Jur9616820}. Although the Kondo peaks in $A_{is}(\w)$s seem not high in \Fig{fig2}(b), their effects are quite significant on both of the magnetic and transport properties. For the latter, the nonequilibrium Kondo resonance assists the electrons transport, which is characterized by the low-temperature current enhancement shown in \Fig{fig3}(c) , when the FM phase dominates at $V>0.25$ mV.

In summary, we have theoretically reported a robust ferromagnetic phase under nonequilibrium conditions in series-coupled double quantum dots  by nonperturbatively solving the Anderson two-impurity model. The ferromagnetic exchange interaction origins from the passive parallel spin arrangement caused by the Pauli exclusion principle during the electrons transport. The ferromagnetic phase can conduce to understand the Heisenberg's initial idea of ferromagnetic order. In addition, it also predicts a convenient way to  internally control spin states without magnetic field.

The support from the NSFC (Grant No.~11374363) and Research Funds of Renmin University of China (No.~11XNJ026) is gratefully appreciated.

\bibliographystyle{unsrt}

\end{document}